\documentclass[a4paper,11pt]{article}
\pdfoutput=1 % if your are submitting a pdflatex (i.e. if you have
\usepackage{jinstpub} % for details on the use of the package, please
\usepackage{float}
\usepackage{siunitx}
\DeclareSIUnit\parsec{pc}
\DeclareSIUnit\lightyear{ly}
\DeclareSIUnit\DAY{day}
\DeclareSIUnit\YEAR{year}
\DeclareSIUnit\speedOfLight{c}
\usepackage[inline]{enumitem}

\title{\boldmath ACHINOS: A Multi-Anode Read-Out for Position Reconstruction and Tracking with Spherical Proportional Counters}

\author[]{I. Katsioulas,}
\author[1]{P. Knights,\note{Corresponding author.}}
\author[]{I. Manthos,}
\author[]{J. Matthews,}
\author[]{K. Nikolopoulos,}
\author[]{T. Neep,}
\author[]{and R. Ward}

\affiliation{School of Physics and Astronomy, University of Birmingham, Birmingham, B15 2TT, United Kingdom}

\emailAdd{p.r.knights@bham.ac.uk}

\abstract{The spherical proportional counter is a versatile gaseous detector with physics applications ranging from rare event searches to fast neutron spectroscopy. In its simplest form, the detector operates with a single channel read-out, and uses pulse-shape information to reconstruct the interaction radius, which is used for background discrimination and target volume definition. Recent developments in the read-out instrumentation have enabled the use of a multi-anode read-out structure, ACHINOS. The multiple anodes provide information about the interaction position which, coupled with the radial information, can be used to reconstruct an ionisation track. This ability has implications for several applications of the detector, for example, background discrimination in rare event searches.  }


\proceeding{12$^{\text{th}}$ International Conference on Position Sensitive Detectors - PSD12\\
  12-17 September 2021\\
  University of Birmingham, UK}

\begin{document}
\maketitle
\flushbottom

\section{Introduction}
\begin{figure}[H]                 
	                    \centering
	      \includegraphics[width=0.34\textwidth]{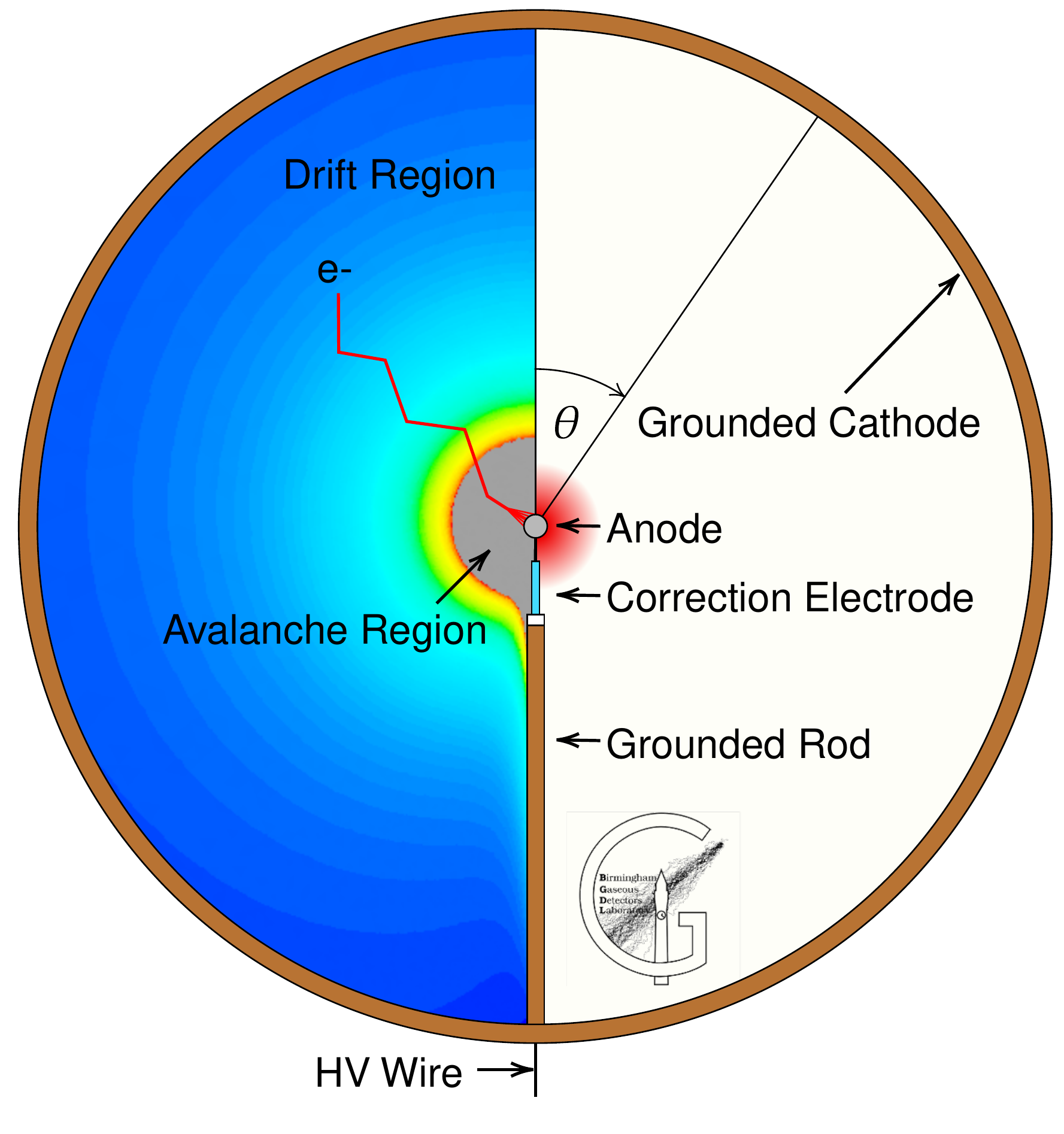}	
	        \includegraphics[width=0.4\textwidth]{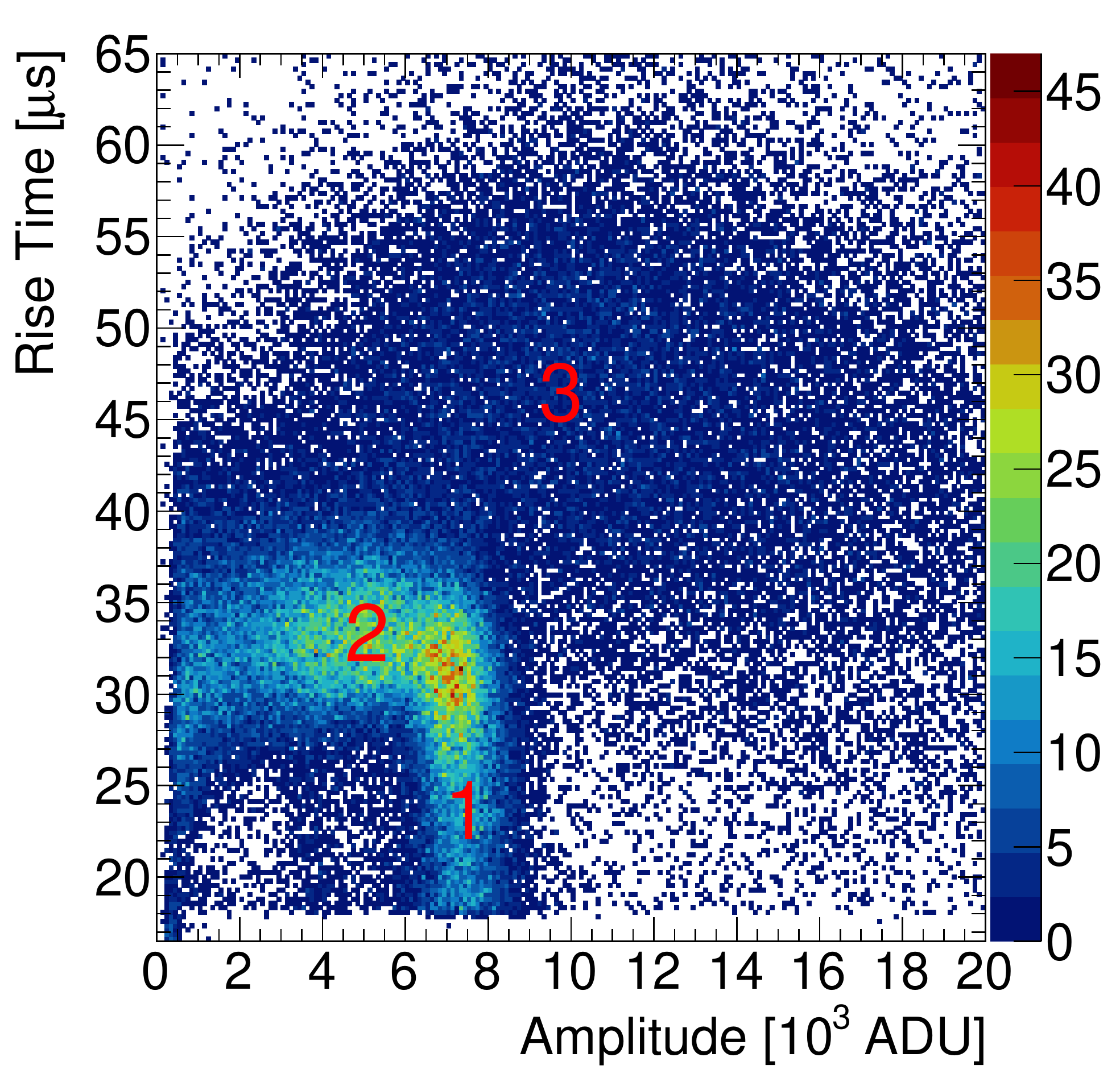}
	        \caption{Left: Schematic showing the spherical proportional counter principle of operation~\cite{Katsioulas:2019sui}. Right: Pulse rise time versus amplitude in an $30\;\si{\centi\meter}$-diameter spherical proportional counter filled with $1.3\;\si{\bar}$ He:Ar:CH$_{4}$ ($51.7\%$:$46\%$:$2.3\%$) and $^{55}$Fe source inside detector. The three distinct populations are (1) $5.9\;\si{\kilo\eV}$ photons interacting in the gas volume, (2) interactions near the cathode, and (3) cosmic muon interactions. \label{fig:psd}}	      
	    \end{figure}

	      The Spherical Proportional Counter~\cite{Giomataris:2008ap} is a simple and robust gaseous detector, well suited to various applications ranging from fast neutron spectroscopy~\cite{Giomataris:2021fwv} to direct Dark Matter (DM) searches~\cite{Arnaud:2017bjh}. Shown in Figure~\ref{fig:psd}, it comprises a grounded spherical cathode, $\mathcal{O}(1\;\si{\meter})$ in diameter, with a read-out electrode in the centre at high voltage. In the simplest case, the read-out electrode is an $\mathcal{O}(\si{\milli\meter})$ in diameter spherical anode. The electric field in the volume varies as $1/r^{2}$, which naturally divides the volume into a drift and avalanche region. Ionisation electrons generated in the gas volume drift toward the anode until approximately $100\;\si{\micro\meter}$ from its surface where the field becomes sufficiently large for charge amplification.	      
	      The key advantages of the detector are:
              %\vspace{-0.4cm}
              \begin{enumerate*}[label=(\roman*)]
              \item{Low capacitance ($\mathcal{O}(0.1\;\si{\pico\farad})$), independent of cathode radius;}
              \item{Single- or few-channel read-out;}
              \item{Stable operation at high gains;}
              \item{Target volume definition (fiduclialisation) and background suppression from pulse-shape characteristics; and }
              \item{Variable gas mixture and pressure.}
              \end{enumerate*}
                        
\section{Radius Reconstruction and Pulse Shape Discrimination}
	      The radial electric field provides a means of Pulse-Shape Discrimination (PSD). Electrons originating from large radii undergo more diffusion while drifting to the anode, resulting in a greater spread in their arrival times. This translates to an increase in measured pulse rise time. Spatially-extended ionisation electron distributions also result a greater spread in anode arrival times. Thus, PSD can be used to identify different interaction types and sources.
	      
	      Figure~\ref{fig:psd} shows data collected with a spherical proportional counter furnished with a single-anode read-out and an $^{55}$Fe source inside, providing $5.9\;\si{\kilo\eV}$ x-rays. Three distinct populations are visible: (1) x-rays interacting in the gas volume at increasing radius for increasing rise time; (2) interactions near the cathode; and (3) cosmic muon interactions, leaving extended ionisation tracks in the detector. PSD allows these three interaction types to be distinguished. Furthermore, the radial information inferred from the rise time allows fiducialisation of the detector.
 
   \section{A Multi-Anode Read-Out: ACHINOS}	 

\begin{figure}[!h]  
	                    \centering
	      \includegraphics[width=0.375\textwidth, trim = 9cm 0cm 9cm 4cm, clip]{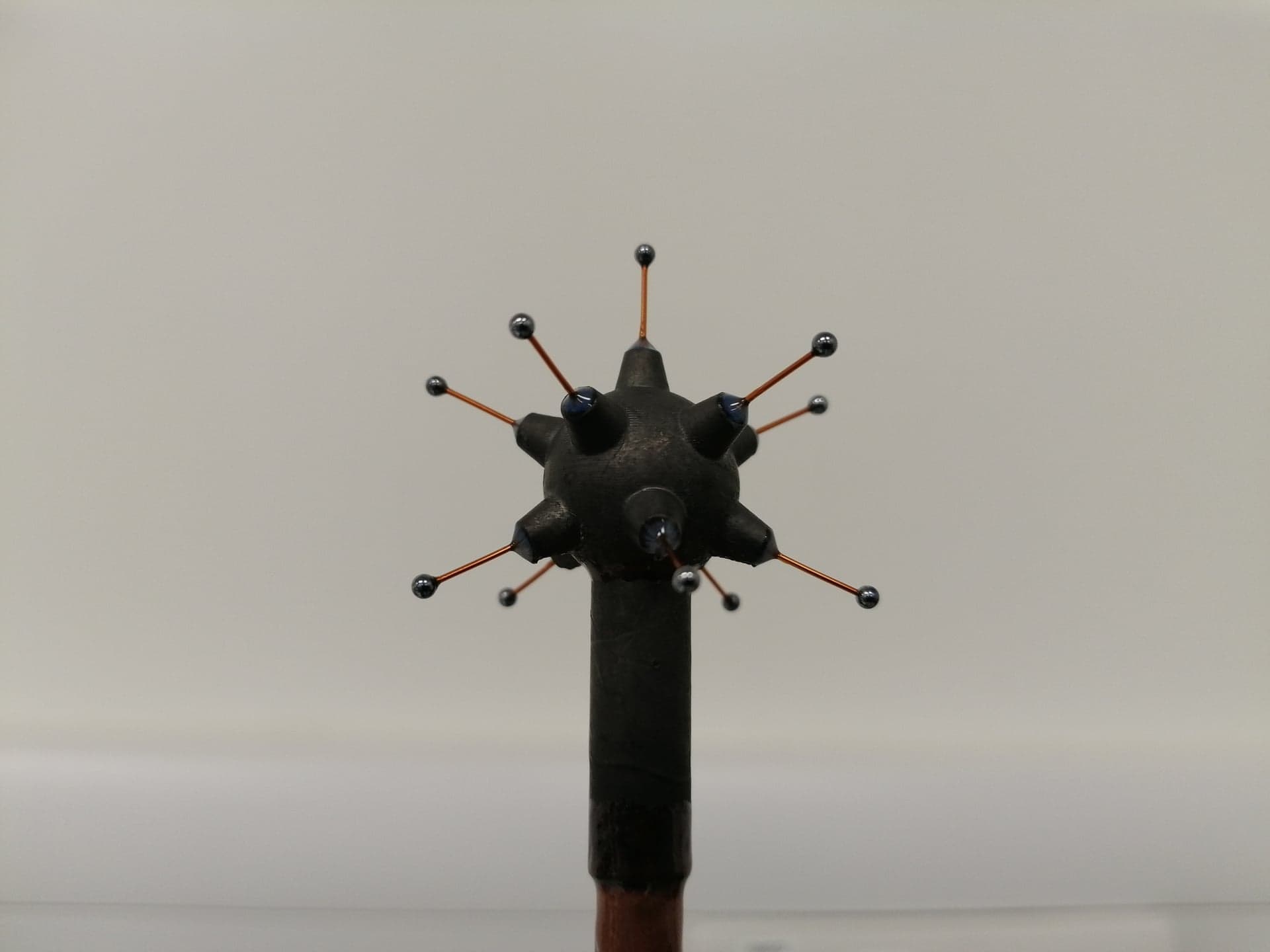}
	      \includegraphics[width=0.4\textwidth]{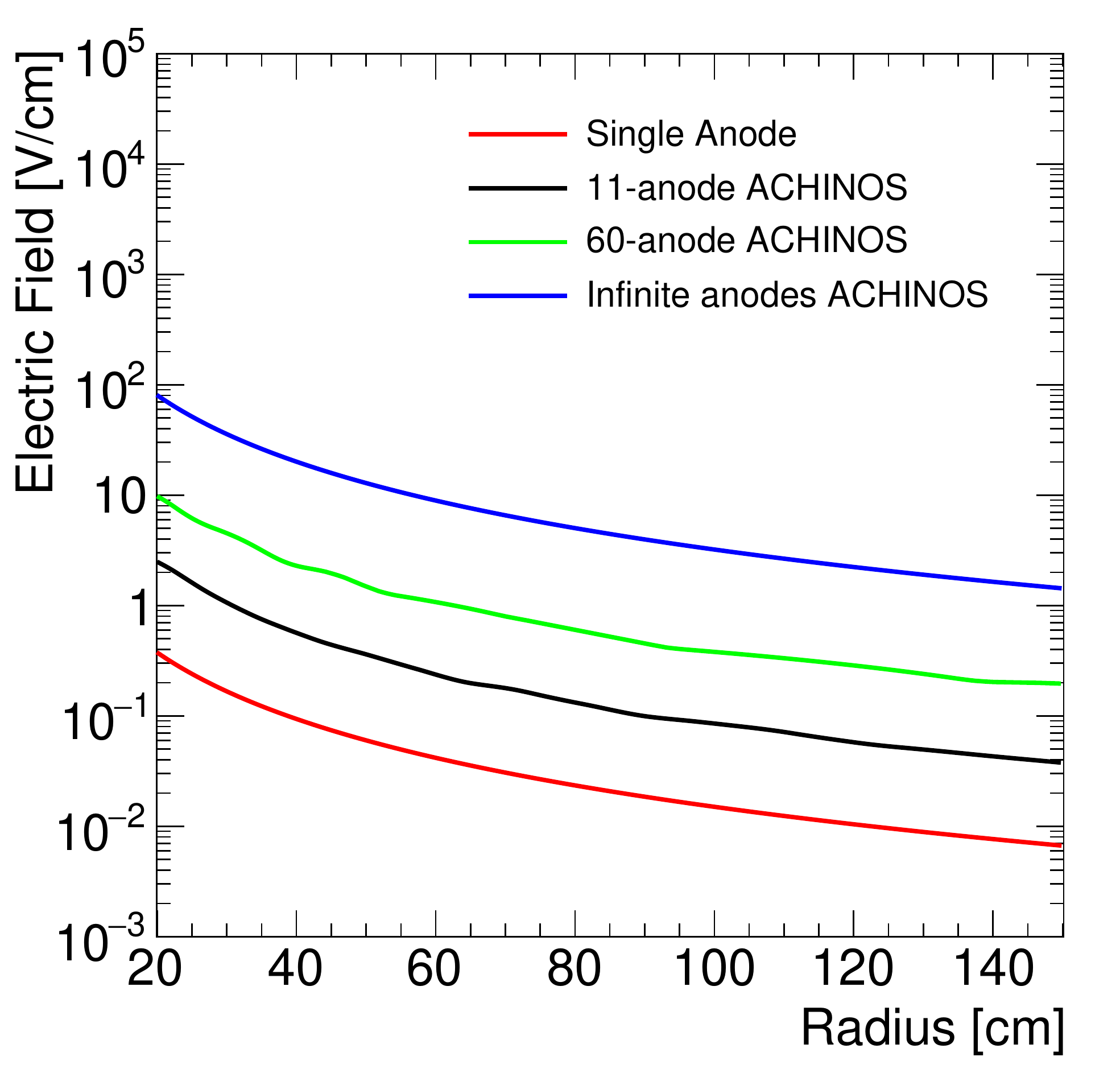}
	\caption{Left: ACHINOS with eleven $1\;\si{\milli\meter}$ anodes and using Diamond-Like Carbon coated 3D-printed central electrode. Right: Electric field magnitude comparison between read-out systems.\label{fig:achinos1}}	      
	    \end{figure}

    Conventional read-out uses a single spherical anode at the detector's centre~\cite{Katsioulas:2018pyh}. In this case, the drift and avalanche fields are coupled, both determined by the anode size and voltage. This limits the detector size, operating pressure and stability.  ACHINOS, shown in Figure~\ref{fig:achinos1}, overcomes this by using multiple anodes at a fixed radius from the detector centre, supported by a central electrode~\cite{Giganon:2017isb, Giomataris:2020rna}. This decouples the avalanche field, which is determined by the size and voltage of each individual anode, and the drift field, determined by the collective electric field of the anodes. 
                 
\begin{figure}[!h]%{r}{0.53\textwidth}
	        \centering
\includegraphics[width=0.65\textwidth, trim = 3cm 0cm 3cm 0cm, clip]{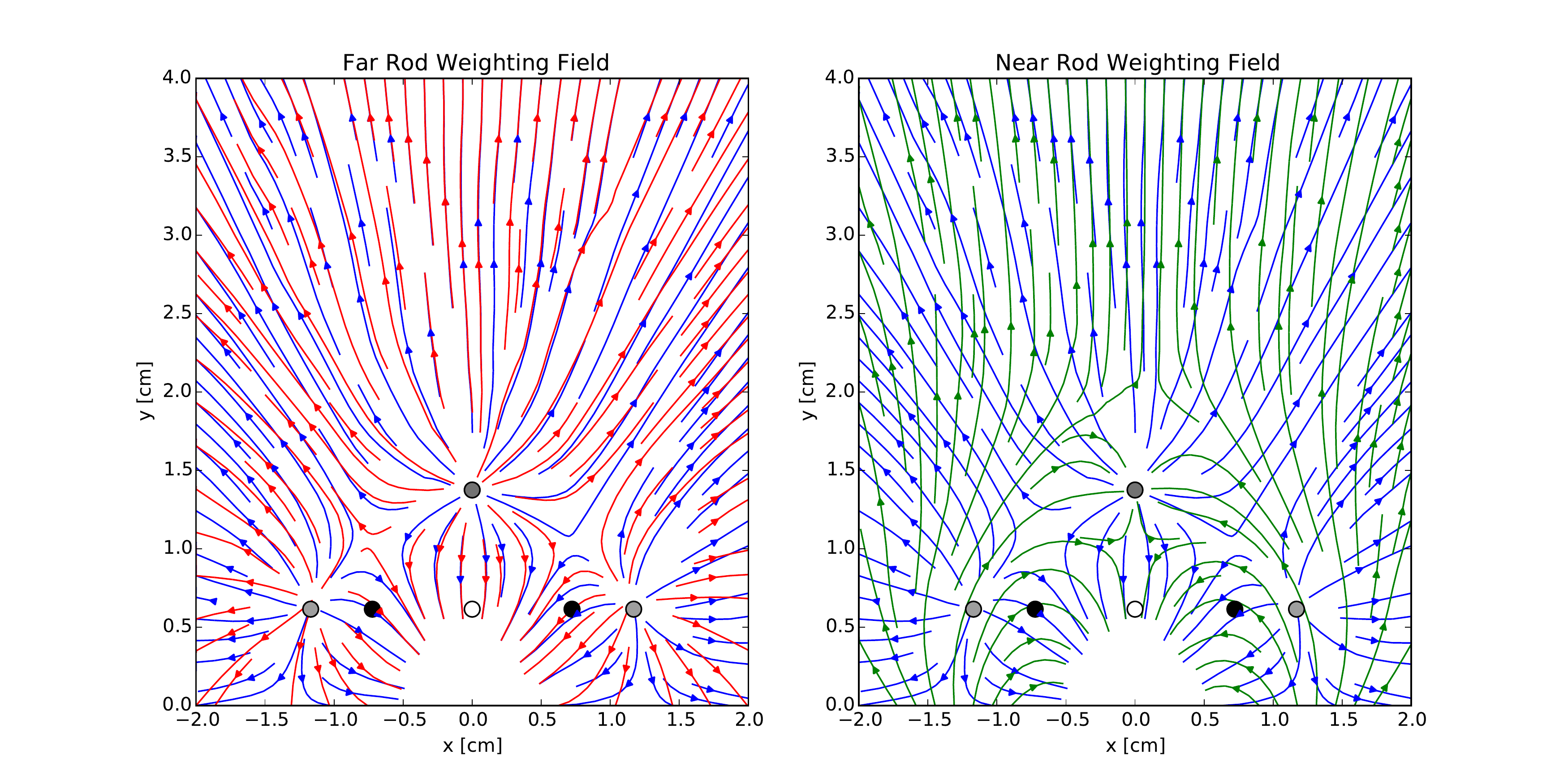} 	       
	        \includegraphics[width=0.32\textwidth]{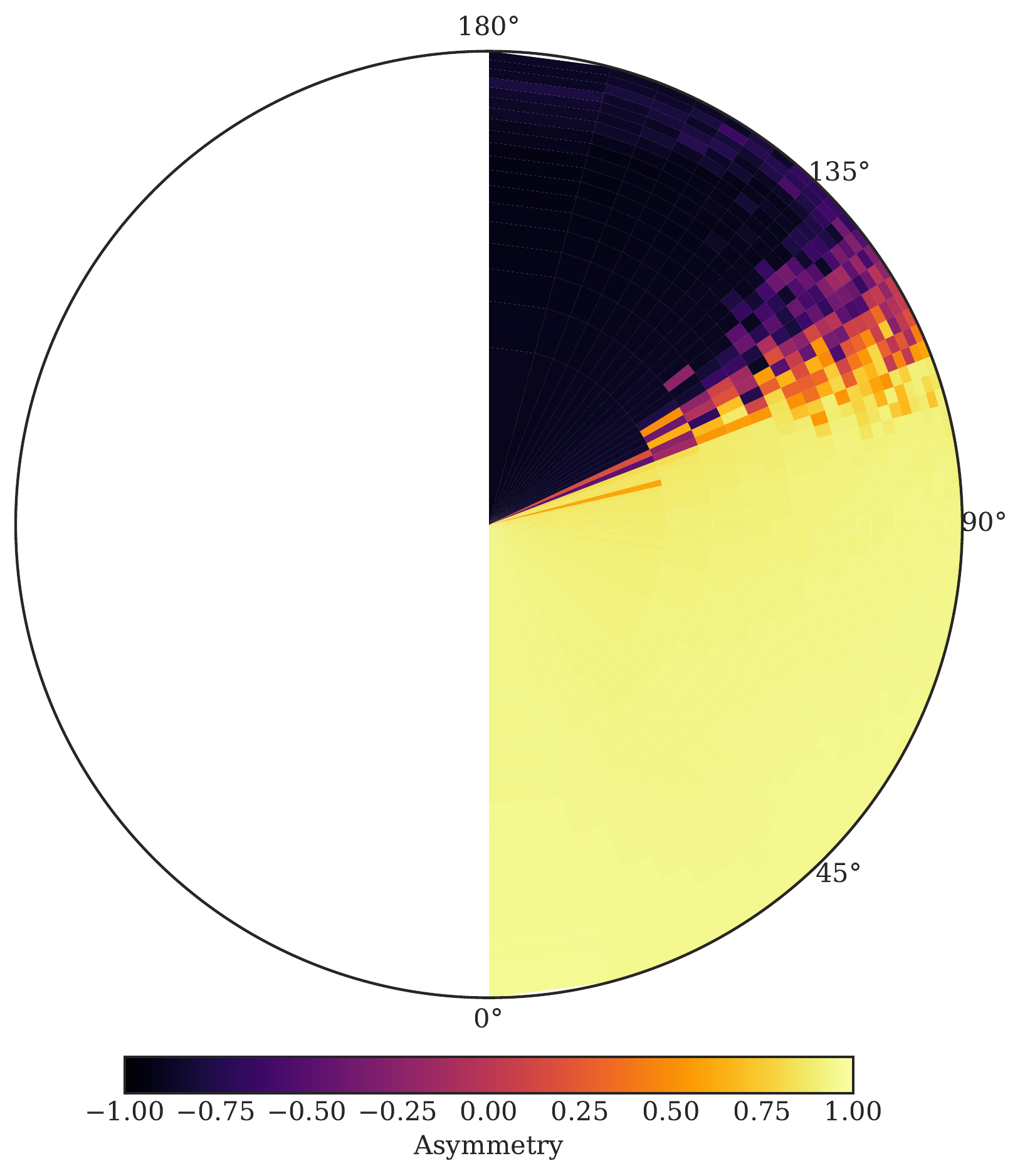}
	        \caption{Left: Weighting fields of the Near (green) and Far (red) electrodes in the vicinity of the Far anodes (circles). The shading of the anodes indicates their depth in the plane (z position), which passes through the anode with the greatest y position. Lighter (darker) shading indicates positions out of (into) the plane. For the anodes in or near the plane (grey shading), the Near weighting field is antiparallel to the electric field (blue) in the vicinity of the anode. 
  	        Right: 2-dimensional histogram of the simulated signal amplitude asymmetry between the Near and Far channels , $A = (F-N)/(F+N)$, plotted for the initial particle location.\label{fig:simAchinos}}
	      \end{figure}

              Extensive simulation~\cite{Katsioulas:2019sui} and experimental testing have been performed to study a single-channel read-out, 11-anode ACHINOS~\cite{Giomataris:2020rna}. It was found that the 5 anodes nearest the rod (`Near') have a higher gain than the 6 further anodes (`Far'), due to their proximity to the grounded rod. This can be corrected for by applying different voltages to Near and Far anodes. Furthermore, by separating the read-out into two channels, spatial information regarding the initial interaction becomes available.  Due to the configuration of the weighting field for each electrode, presented in Figure~\ref{fig:simAchinos}, the signal generated in one electrode when all the primary electrons arrive to the other electrode will be inverted relative to the case where all electrons arrive to that anode. Thus, using amplitude asymmetry, the fraction of primary electrons arriving to each electrode can be deduced and, so, the detector region where the interaction occurred, as shown in Figure~\ref{fig:simAchinos}.
              
In the future, individual anode read-out of ACHINOS will provide higher granularity spatial information. Larger spherical proportional counters and higher pressure operation require ACHINOS with more anodes, as demonstrated in Figure~\ref{fig:achinos1}. The feasibility of a 60-anode ACHINOS is under investigation. Moreover, additional anodes have potential to improve the position resolution.

	    \section{Summary}
	   Instrumentation developments for the spherical proportional counter are enabling the operation of larger detectors and at higher pressures, which are essential requirements for ongoing and future applications. While the single-anode read-out provides radial information and track-like ionisation discrimination, ACHINOS has the potential for interaction position and track reconstruction in the detector by reading out more than one electrode from the anodes. Future implementations with individual anode read-out, as well as more anodes, will enable track and position reconstruction.

\bibliographystyle{ieeetr}
\bibliography{mybib} 
\end{document}